\documentstyle[aps,prl,multicol,graphicx]{revtex}

\begin{document}
\draft
\title{$\mathbf{q}$-dependence of low-frequency Raman scattering in silica glass}
\author{N.\thinspace V.~Surovtsev\cite{sur}, J.~Wiedersich, V.\thinspace N.~Novikov%
\cite{sur}, E.~R\"{o}ssler}
\address{Physikalisches Institut, Universit\"{a}t Bayreuth, D-95440 Bayreuth, Germany}
\author{E. Duval}
\address{LPCML, Universit\'{e} Lyon I, F-69622 Villeurbanne C\'{e}dex, France}
%\date{January 20, 1999: submitted to Physical Review Letters, accepted}
\maketitle
 
\begin{abstract}
Accurate measurements of the dependence of low-frequency Raman scattering
on the scattering angle were performed in two silica glasses. By a 
comparison of spectra measured at a large scattering angle (close to back
scattering) and a small one (close to forward scattering), we for the
first time 
observed a $\mathbf{q}$-dependence of the low-frequency light scattering
in glasses. From the magnitude of the effect, the vibration
correlation length is estimated and compared with results from the
picosecond optical technique; a reasonable agreement is found.
\end{abstract}

\begin{multicols}{2}

Vibrational excitations in glasses show two peculiarities: an excess density
of vibrational states with respect to the Debye one in the THz region (the
boson peak) and a strong scattering of phonons at frequencies above around
100\thinspace GHz. These peculiarities appear at acoustic wavelengths of the
order of nanometers and demonstrate themselves, e.\,g., in the anomalous
thermal conductivity and specific heat at low temperatures 
(5$-$10\,K) \cite{Poh}, and in inelastic neutron and light scattering spectra 
\cite{Buch,Jac}. Many works over the past two decades have been devoted to the
study of these topics, but both the theoretical understanding and the
experimental investigation of vibrational properties of glasses are far from
being complete. New experimental techniques, like inelastic X-ray scattering
(IXS) and small-angle inelastic neutron scattering (INS) open additional
possibilities to study acoustic spectra in the Brillouin scattering regime
and in the frequency range of the boson peak \cite{IXS1,IXS2,IXF}. Together
with the picosecond optical technique (POT) \cite{POT1}, the results of these
methods encourage further investigations of the dynamic and spatial
properties of acoustic excitations in the range of 100$-$3000\thinspace GHz
\cite{Buch2,Vach2}. However, the analysis of the results has been the
subject of some controversy, that cannot be resolved at present due to the
limited range of spectral resolution of the devices (see, e.\,g., Ref. \cite{cIXS}). 
Therefore, it would be very useful to obtain additional information
about this subject from other experimental methods.

Traditionally, low-frequency Raman scattering is applied to study spectral
properties of vibrational excitations in glasses. Usually the momentum
dependence of the scattering ($\mathbf{q}$-dependence) is not considered
\cite{Jac}, because the phonon mean free path in the corresponding frequency
range is much less than the wavelength of the scattered light. As a result,
any $\mathbf{q}$-dependence effect should be very small and difficult to 
observe experimentally.
The $\mathbf{q}$-dependence of light scattering around the
frequency of the boson peak has recently been investigated in SiO$_{2}$
glass at ambient conditions \cite{Q1}; 
no $\mathbf{q}$-dependence was
observed, 
but the accuracy of the measurement allowed an estimation of the upper limit of 
the localisation lenght of vibrations.

In the present Letter
this approach is refined and extended to lower frequencies and lower
temperatures. We present the first observation of a $\mathbf{q}$-dependence
effect for low-frequency Raman scattering in glasses. In order to extend
the available frequency range of the experiments and to demonstrate the
reliability of our results, we use two different, complementary light
scattering techniques: conventional Raman spectroscopy (employing a
monochromator) and an interferometric technique (employing a tandem
Fabry-Perot interferometer). In both cases we use spectrometers with the
best elastic line suppression available in their class.

The Raman scattering experiment on a sample of glassy SiO$_{2}$ (Heralux,
Heraeus, 130\thinspace ppm of OH$^{-}$-groups) was performed using an 
Ar$^{+}$\,laser (514.5\thinspace nm, 900\thinspace mW) and a five-grating
monochromator Z40. The two scattering angles were 20$^{\circ }$ and 
160$^{\circ }$; the respective wave vectors were 
$q_{1}$=0.62$\times 10^{-2}$\thinspace nm$^{-1}$ and 
$q_{2}$=3.5$\times 10^{-2}$\thinspace nm$^{-1}$
with the refractive index of the sample $n$=1.46. An aperture limiter was
placed on the lens in order to collect the scattered light within an angle
of $\sim $14$^{\circ }$. Polarized Raman spectra were measured using
spectral slits of 60\,GHz (2\thinspace cm$^{-1}$) for the spectral interval
420$-$3180\,GHz
(14$-$106\thinspace cm$^{-1}$) and 45\, GHz (1.5 cm$^{-1}$) for the spectral interval
150$-$660\,GHz
(5$-$22\thinspace cm$^{-1}$). The position of the elastic line was checked for
each spectrum with an accuracy of 3\,GHz (0.1\thinspace cm$^{-1}$).

One experiment was performed at ambient conditions. The temperature in the
illuminated volume of the sample obtained from the Stokes/anti-Stokes ratio
was 311.0$\pm $1.7\thinspace K, and within this precision was the same for
experiments in both geometries. 
The temperature 
uncertainty leads to a spectral shape uncertainty lower than 3$\times 10^{-3}$ 
over the measured spectral range. In order to obtain spectra with a
high accuracy, for each scattering angle we recorded 120 and 36 spectra in
the frequency ranges 420$-$2220\thinspace GHz
and 150$-$660\thinspace GHz, respectively.

An Oxford optical helium cryostat was used for the low-temperature
experiments. The incident laser beam entered the cryostat as in a
conventional right angle experiment. We used a small mirror within the
cryostat in order to have the necessary angle of the incident beam relative
to the direction of the scattered light. With this optical scheme we
exclude any Raman scattering contribution from cryostat windows that would 
be difficult to avoid when one uses the same window (or two parallel ones) for
the incident laser beam and for the scattered light in the case of a
monochromator with slits. The temperature in the illuminated volume of the
sample was determined by the Stokes/anti-Stokes ratio to be 
33.0$\pm $0.5\thinspace K. 
72 and 40 spectra are accumulated for each scattering angle
in the frequency ranges 150$-$660\,GHz
and 420$-$3180\,GHz,
respectively. We include low-frequency Raman scattering spectra of
Heralux and Suprasil 
measured in previous experiments
as right-angle Raman scattering, using the same
laser and monochromator. The temperature of the illuminated part of the
sample was 7\thinspace K in the case of Heralux and 4\thinspace K in the
case of Suprasil as obtained from the Stokes/anti-Stokes ratio.

\begin{figure}[!bth]
\begin{center}
\includegraphics*[height=2in]{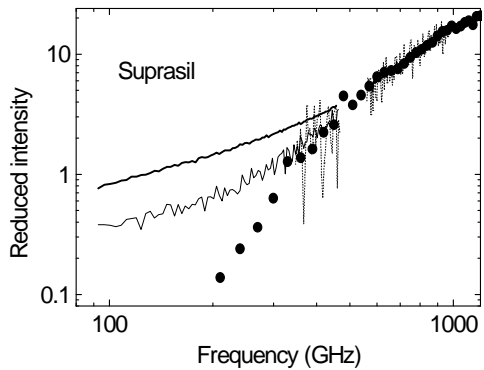}
\caption{ Reduced light scattering spectra of Suprasil in the forward
scattering experiment measured by the Sandercock tandem at $T$=300\,K (thick
line), 45\,K (thin solid line), 6\,K (dotted line). Symbols corresponds to
the right-angle monochromator spectrum at $T$=4\,K. }
\label{spectrum}
\end{center}
\end{figure}
Inelastic light scattering spectra of a sample of Sup\-ra\-sil 300 (synthetic
silica, Heraeus, $<$1\thinspace ppm of OH$^{-}$-groups) were obtained at
temperatures 300, 45 and 6\thinspace K (employing a CryoVac dynamic Helium
cryostat) using a six-pass Sandercock tandem Fabry-Perot interferometer 
\cite{sand} and an Ar$^{+}$\,laser (514.5\thinspace nm, 600\thinspace mW). The
scattered light was recorded with no selection of polarization; the
polarization of the incident beam was perpendicular to the scattering plane
leading to a domination of the polarized inelastic light scattering in this
experiment due to the low depolarization ratio of silica glass ($\sim $0.3
at low frequencies). In this experiment either of two anti-parallel beams of equal
intensity was focused onto the same volume of the sample. The 
direction of the scattered light corresponds to scattering angles of 
7$^{\circ }$ and 173$^{\circ }$ 
for the respective beams 
($q_{1}$=0.21$\times 10^{-2}$\,nm and $q_{2}$=3.6$\times 10^{-2}$\,nm),
and was collected within an angle of $\sim 16^{\circ}$. 
We used a free spectral range of 500\thinspace GHz over two spectral ranges
on either side of the elastic line. The position of the elastic line was
constantly kept aligned within about 4\thinspace GHz. 
The entrance and exit pinholes of the spectrometer were 450 and
700\thinspace $\mu $m, respectively,
in order to suppress the transmission of higher orders of the tandem 
interferometer in combination with a prism
(the tandem
transmits every 20th order of a free spectral range; without suitable
suppression this could significantly disturb a broad spectrum) \cite{Sander}. 
The experimentally determined finesse was better than 100.

To further validate the absence of possible contributions from the
instrumental tail of the elastic line or from higher transmission orders of
the tandem, we measured spectra for both geometries at a low temperature, 
$T$=6\thinspace K. 
At such a very low temperature the anti-Stokes part of the
spectrum should be almost zero because the scattering intensity of the
anti-Stokes spectral side is proportional to the Bose factor which is very
small for this temperature (in the frequency range where this factor is not
very small for this temperature, $\nu $$<$200\,GHz, the signal itself is 
very low). 
Indeed, the anti-Stokes part of the spectrum at this temperature shows no 
deviations from the dark count level of 2.47\,cts/s of our detector, 
demonstrating the absence of contributions of higher orders or from the
elastic line.
A quantitative estimation limits any 
systematic deviations to less than about 0.03 cts/s for the
frequency range down to 100 GHz. This value is three orders of magnitude
less than the level of the signal at $T$ = 300 K which was $\geq $ 50 cts/s.
At $T$ = 45 K the level of the signal varies from 1.16 cts/s at 100
GHz to 10 cts/s at 1000 GHz. Thus, one can expect that a possible distortion
of the experimental spectra is not higher than 0.3\% for 1000 GHz and 3\%
for 200 GHz at 45 K.
 
A low-frequency light scattering spectrum of glasses consists of two
contributions---the vibrational spectrum and a quasielastic one (QES); the
latter increases faster than the Bose factor as temperature increases and
dominates at low frequencies. Fig.~\ref{spectrum} shows the reduced spectra
(i.\,e., $I/(n(\nu )+1)$ of the Stokes side, where $n(\nu )$ is the Bose
factor) of the Suprasil sample measured by the Sandercock tandem at $T$=300,
45 and 6 \thinspace K in the forward scattering geometry; the right-angle
monochromator spectrum recorded at 4\thinspace K is also included. At
frequencies above some 600\thinspace GHz the vibrational contribution
dominates and the reduced spectrum does not depend on temperature. At lower
frequencies it increases with temperature, since QES becomes significant. At
$T$=45\thinspace K the vibrational spectrum dominates at least down to
300\thinspace GHz.
\begin{figure}[t!]
\begin{center}
\includegraphics*[height=1.7in]{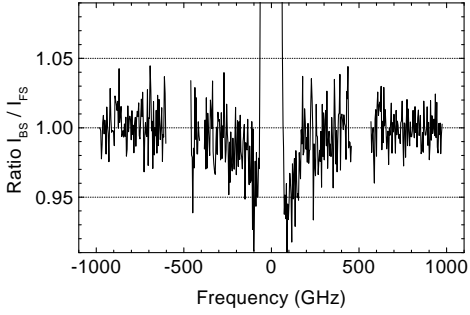}
\caption{ Ratio of back- and forward-scattering spectra 
($I_{\mathrm{BS}}$ and $I_{\mathrm{FS}}$) of Suprasil at $T$=300\,K. }
\label{r300}
\end{center}
\end{figure}

\begin{figure}[!bth]
\begin{center}
\includegraphics*[height=2.9in]{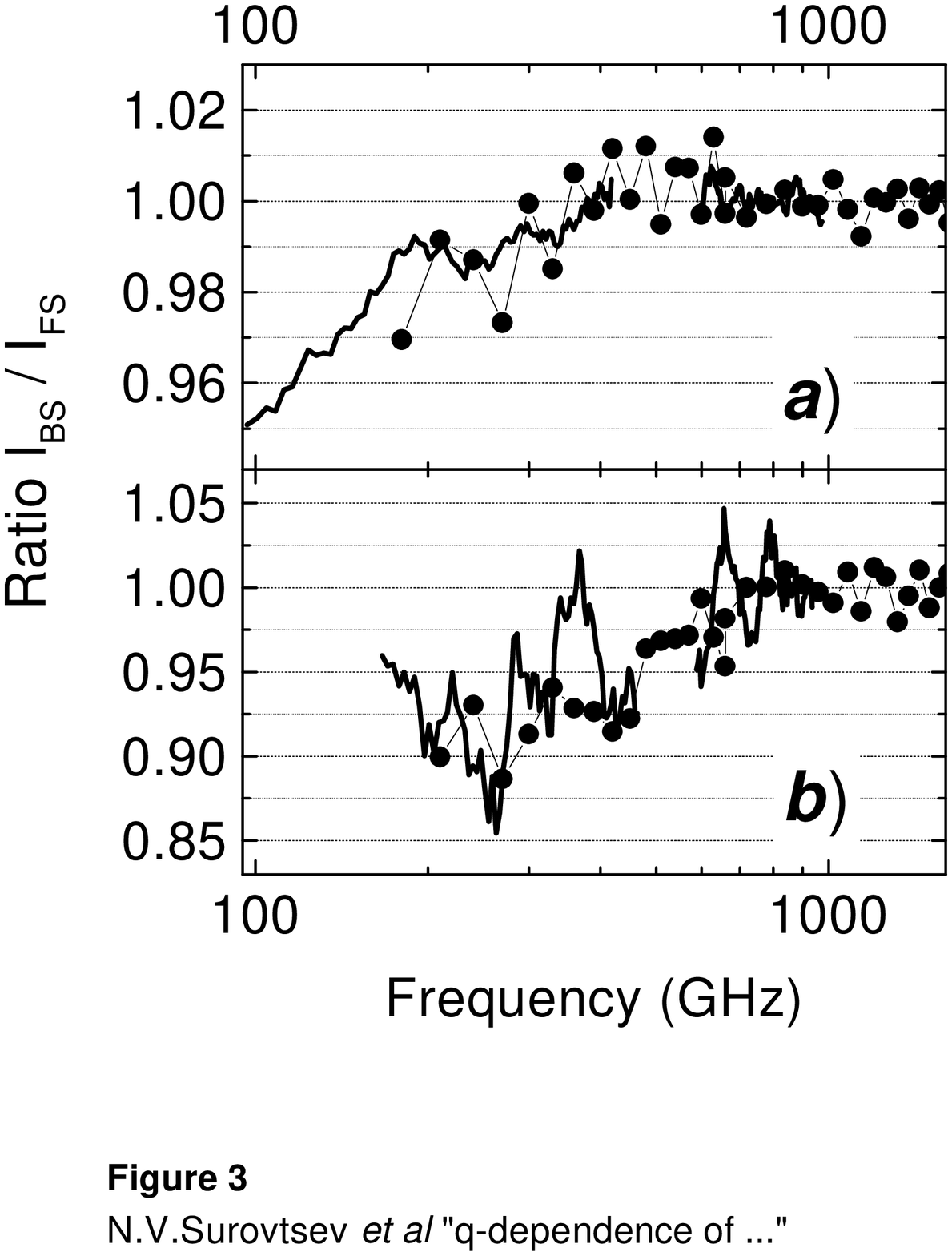}
\caption{ Ratio of back- and forward scattering spectra of Suprasil (solid
line) and Heralux (circles) obtained a) at room temperature; b) at $T$=33\,K
for Heralux and $T$=45\,K for Suprasil. }
\label{ratio}
\end{center}
\end{figure}

The ratio of the back- and forward-scattering spectra, $R$, of Suprasil
measured with the Sandercock tandem at $T$=300\thinspace K is shown in 
Fig.~\ref{r300} for the Stokes and anti-Stokes sides. This ratio is a constant at
high frequencies; we fix its value as 1 since it 
is constant up to the optical modes \cite{Q1}. The figure shows that the
ratio deviates from a constant in the frequency range
100$-$300\thinspace GHz, which is a clear demonstration of 
the $\mathbf{q}$-dependence
of the light scattering. To improve our statistics, in Fig.~\ref{ratio} 
(a) we show this ratio $R$ averaged over the Stokes and anti-Stokes
sides and smoothed by adjacent averaging over 10 points. Circles in 
Fig.~\ref{ratio} (a) correspond to the result obtained for Heralux by Raman
scattering. We note that in spite of a very weak $\mathbf{q}$-dependence
(about a few percent) and different experimental techniques being used in
these two cases, we find an excellent agreement between the two curves.
This is a strong evidence for the reliability of the $\mathbf{q}$-dependence 
effect found in our experiments. Fig.~\ref{ratio} (b) shows the
back- to forward scattering ratio $R$ obtained for Suprasil at 
$T$=45\thinspace K and for Heralux at $T$=33\thinspace K. Again a clear
indication of the $\mathbf{q}$-dependence of the light scattering is
found. The precision of these curves is lower than in Fig.~\ref{ratio} (a)
due to a decrease of the signal at low temperatures; however, the
magnitude of the $\mathbf{q}$-dependence effect is larger. Again, there
is a good agreement between the results obtained by the different
experimental techniques. A comparison of these results for the different
temperatures with Fig.~\ref{spectrum} leads to the conclusion that the 
$\mathbf{q}$-dependence effect increases with a decrease of the
quasi-elastic contribution with respect to the vibrational contribution. The
deviation of the $\mathbf{q}$-dependence from a constant is stronger for
lower frequencies; this corresponds to an increase of the vibration
correlation length with decreasing frequency, 
as will be explained in the following.

Let us turn to the interpretation of the experimental data. The 
$\mathbf{q}$-dependence of the inelastic light scattering intensity in the
acoustic region is determined by the equation \cite{Q1}
\begin{equation}
I(q,\nu )\propto F_{\nu }(q)\left\langle \left| \nabla u_{\nu }(0)\right|
^{2}\right\rangle g(\nu )  \label{e0},
\end{equation}
where $g(\nu )$ is the vibrational density of states, $u_{\nu }(r)$ is the
amplitude of a vibration of frequency $\nu $ and $F_{\nu }(q)$ is the
spatial Fourier transform of the vibration correlation function
\begin{equation}
F(r)=\langle \nabla u_{\omega }(\vec{r})\nabla u_{\omega }(0)\rangle
/\langle \left| \nabla u_{\omega }(0)\right| ^{2}\rangle .  \label{e1}
\end{equation}
At small $q$, $F_{\nu }(q)\propto 1-b(ql_{\nu })^{2}$ where $b$ is a
constant which depends on a particular form of $F(r)$ and $l_{\nu }$ is the
vibration correlation length \cite{Q1}. To evaluate the 
$\mathbf{q}$-dependence of the experimental spectra, we use the ratio of $I(q,\nu )$
measured at the larger wavevector, $q_{2}$ (close to back scattering), to
that at the small scattering angle corresponding to the wavevector $q_{1}$\cite{qexp}:
\begin{equation}
R(q_{1},q_{2},\nu )=I(q_{2},\nu )/I(q_{1},\nu )\cong
1-b(q_{2}^{2}-q_{1}^{2})l_{\nu }^{2}.  \label{e2}
\end{equation}
In Ref. \cite{Q1} it was shown that typical correlation functions (e.\,g.,
Gaussian or exponential) lead to a parameter $b$ in the range from 0.1
to 2. In the case of attenuated plane waves the respective expression for 
$F_{\nu }(q)$ is more complicated and will be considered in a forthcoming
paper. 
For a rough estimate we 
follow Shuker and Gammon \cite{sg} and 
approximate $F(r)$ by a step function: 
 $F(r)={\mathtt const}$ at $r<l_{\nu} $ 
and $F(r)=0$ at $r>l_{\nu }$.
This leads to a value of $b=0.1$ (see, e.\,g., Ref.~\cite{Q1}).

The expression (\ref{e2}) corresponds to the vibrational part of the
spectrum. At low frequencies (below 300\,GHz at 45\,K and 600\thinspace GHz at
300\thinspace K), the QES gives a significant contribution; its 
$\mathbf{q}$-dependence is not known
and in principle may be different from that of
vibrations as a function of frequency. So, we first consider the
part of our spectra which is dominated by vibrations. Our data at 45 and 33\,K
show a clear $\mathbf{q}$-dependence in the region between 300 and 
600\,GHz. 
To estimate the correlation length $l_{\nu}$ from the observed 
$\mathbf{q}$-dependence by Eq.~\ref{e2}, we use our value of $R$ from 
Fig.~\ref{ratio} at, e.\,g., 
320\, GHz. This leads to $l_{\nu}$=27\,nm at 33\,K. 

It is interesting to compare $l_{\nu }$ with the mean free path $L_{\nu }$
of the acoustic vibrations obtained by POT \cite{POT1}, because it is
natural to expect that $L_{\nu }$ and $l_{\nu }$ are interrelated. In 
\cite{POT1} an estimate $L_{\nu }$=24\thinspace nm 
was obtained for $L_{\nu }$ in
silica at 342\thinspace GHz and room temperature. 
Therefore we need
to know $l_{\nu }$ at 300\,K, 
where one should separate the vibrational contribution from that of QES
in the data of $R$.
We assume that
within our accuracy the 
$\mathbf{q}$-dependence of QES is negligible. 
This is in accordance with the model of QES developed, e.\,g., in Ref.\ 
\cite{noqqes}.
In these papers it is argued that the $\mathbf{q}$-dependence of  QES is
the same as that of the boson peak. 
However, at the frequencies of the boson peak, 
as it was already shown in Ref. \cite{Q1}, a 
$\mathbf{q}$-dependence is absent with a precision of 0.3\%, i.\,e.\ it is
unobservable with our accuracy. 
The idea that QES is related to relaxational modes that are localized 
on a short range 
leads to the same conclusion. Therefore, assuming that 
$R_{{\rm {relax}}}$=1, it is easy to show that 
\begin{equation}
R(q_{1},q_{2},\nu )\cong 1-\eta b(q_{2}^{2}-q_{1}^{2})l_{\nu }^{2},
\label{e3}
\end{equation}
where $\eta =I_{{\rm vib}}/I$. Now, using Eq.\ (\ref{e3}), taking $R$ from
Fig.~\ref{ratio} and $\eta $ from Fig.~\ref{spectrum} (with the spectrum at 
$T$=4\thinspace K as the vibrational contribution $I_{{\rm vib}}$) one
obtains $l_{\nu }$= 14\thinspace nm at $T$=300\thinspace K and $\nu =$ 320
GHz. This estimate is made for a particular choice of the parameter $b$. For
different choices of $b$ the magnitude of $l_{\nu }$ will change 
as $\sqrt{b}$. The fact that the values of $l_{\nu }$ and $L_{\nu }$
have the same order of magnitude may be an additional evidence that the
mechanism of the $\mathbf{q}$-dependence we use in this paper is correct.

We note, that at $T$=33\thinspace K the correlation
length is by a factor of 2 bigger than at room temperature. Our result
agrees with the conclusion of Ref.\ \cite{POT1} that at least up to
frequencies of 400\thinspace GHz the vibrational mean free path decreases with
increasing temperature up to 100 K. The conclusion of Ref.\ \cite{POT1} was
based on a comparison of the data for the phonon mean free path from the
POT \cite{POT1}, the thermal conductivity of SiO$_{2}$ \cite{TC} and from
the tunnel junction technique \cite{TJ} applied at different temperatures.
Our result leads to the same conclusion for $l_{\nu }$ (independent of the
choice of $b$), but is obtained by a single experimental technique. It is
interesting to note that the ratio of $l_{\nu }$ at $T$=300\thinspace K to
that at 33\thinspace K in our case (a factor of 2 for $\nu $=320\thinspace
GHz) is similar to that obtained by Brillouin scattering at 
$\nu $=35\thinspace GHz \cite{BL}; 
this means that the dominating mechanism of the
phonon attenuation may be the same at both frequencies at $T$$>$30\,K.

In conclusion, we present the first observation of a $\mathbf{q}$-dependence
of the low-frequency (0.1-1\thinspace THz) light scattering spectra of
silica glasses. This effect provides information on the spatial properties
of THz dynamics in glasses. We demonstrate that the estimate of the
vibration correlation length found from the $\mathbf{q}$-dependence of
light scattering is in a reasonable agreement with the vibrational mean free
path found by POT.

Helpful discussions with Prof.\ A.\,P.~Sokolov and 
Prof.\ V.\,K.\ Malinovsky are
appreciated. N.\,V.\,S. thanks Universit\'{e} Lyon I and Universit\"{a}t
Bayreuth for hospitality and the support from a grant for young scientists
by the Siberian Branch of the Russian Academy of Sciences. This work has
been supported by SFB 279 of the Deutsche Forschungsgemeinschaft.

\end{multicols}

\begin{references}
\bibitem{qexp} For experimental reasons we had to use slightly different scattering geometries
for the tandem and Raman experiments; 
however, within our precision the value of $(q_2^2-q_1^2)$ is the same for both experiments.
\end{references}
\end{document}